\documentclass{article}
\usepackage{spconf,amsmath,graphicx}
\usepackage{amsmath}
\usepackage{amssymb}
\usepackage{mathtools}
\usepackage{amsthm}
\usepackage{xcolor}
\usepackage{hyperref}
\usepackage{booktabs}
\usepackage{subcaption} %
\usepackage{caption}
\usepackage{multirow}
\usepackage[export]{adjustbox}
\usepackage[normalem]{ulem}
\usepackage{tikz}
\usepackage{pgfplots}
\usepgfplotslibrary{fillbetween}

\definecolor{c4}{HTML}{916cad}
\definecolor{c9}{HTML}{dc7168}

\title{Estimating the completeness of discrete speech units}
\name{Sung-Lin Yeh, Hao Tang}
\address{School of Informatics, University of Edinburgh}
\begin{document}
\maketitle
\begin{abstract}
Representing speech with discrete units has been widely used in
speech codec and speech generation.
However, there are several unverified claims about self-supervised discrete units,
such as disentangling phonetic and speaker information with k-means,
or assuming information loss after k-means.
In this work, we take an information-theoretic perspective to answer
how much information is present (information completeness) and
how much information is accessible (information accessibility),
before and after residual vector quantization.
We show a lower bound for information completeness and estimate
completeness on discretized HuBERT representations after residual vector quantization.
We find that speaker information is sufficiently present in HuBERT discrete units,
and that phonetic information is sufficiently present in the residual,
showing that vector quantization does not achieve disentanglement.
Our results offer a comprehensive assessment on the choice of discrete units,
and suggest that a lot more information in the residual should be mined rather than discarded.
\end{abstract}

\section{Introduction}

Previous work has proposed to use discrete speech units 
as an alternative to a variety of speech tasks,
which offers lower computational and storage costs at some loss in performance.
Of particular interests are discrete units derived from self-supervised speech representations, 
because the representations have demonstrated strong performance in many downstream tasks 
\cite{oord2018representation,baevski2020wav2vec,chung2020vector,hsu2021hubert,chen2022wavlm}.
For example, 
the discrete units, usually realized with k-means on self-supervised representations, 
have been applied to automatic speech recognition \cite{chang2023exploration,yang2024towards},
due to their strong phonetic prominence \cite{hsu2021hubert,wells2022phonetic}.
Recent work has also considered synthesizing speech with discrete speech units 
\cite{polyak2021speech,lakhotia2021generative,kreuk2021textless,yang2024towards,du2024unicats}, 
claiming either that quantization has an disentanglement effect,
or that the speaker identity is lost if not explicitly modeled.
We ask how much information is present (information completeness) and
how much information is accessible (information accessibility) 
before and after vector quantization of speech representations.

Information accessibility is understood as how easy we can extract certain information from the representations,
while information completeness indicates how much information from the original signals
is encoded in the representations.
The accessibility has inspired the development of many probing tasks \cite{yang2021superb,lin2023utility}, 
using accuracy as a proxy to measure how accessible the target information is using a simple classifier
\cite{pimentel2020information}.
However, there is not yet a comprehensive study of the completeness of a representation,
and how it relates to information accessibility.
This question has received considerable attention when it comes to speech generations solely relying on 
discrete speech units from k-means \cite{lakhotia2021generative,lee2021direct,yang2024towards}
or residual vector quantization (RVQ) \cite{zeghidour2021soundstream,borsos2023audiolm,wang2023neural,peng2024voicecraft}, 
in which information is highly likely to lose.

Although recent approaches have proposed to evaluate information completeness of discrete speech units 
on synthesized speech,
the synthesized speech may not faithfully reflect the encoded information 
\cite{polyak2021speech,lakhotia2021generative,zhang2023speechtokenizer}.
For example, 
the synthesizer could hallucinate especially when using generative adversarial networks (GANs) \cite{lucas2019adaptive}.
The additional speech recognition and speaker embedding systems
can amplify the effect of hallucination \cite{frieske2024hallucinations}.
Rather than synthesizing speech, in this work we directly evaluate completeness on the discrete speech units.

To answer how complete a representation is, 
we show a lower bound of mutual information for information completeness through the lens of information theory, 
with which we estimate completeness on discretized HuBERT representations after RVQ.
More specifically, we pose information completeness as minimum distortion
between the representations and associated log Mel spectrograms.
While estimating mutual information is known to be difficult (if not impossible) \cite{mcallester2020formal},
the lower bound has an important interpretation---the amount of information that is at least present in the representations.

We further connect information completeness to information accessibility,
adopting higher-performing probes
to achieve tighter lower bound of mutual information \cite{pimentel2020information}.
We then use the proposed lower bound to examine several design choices and unverified claims
on speech representations and discrete speech units.
For example, Zhang \textit{et al.} \cite{zhang2023speechtokenizer} claim that
``there is significant information redundancy between semantic tokens and acoustic tokens'',
with semantic tokens (a misnomer itself \cite{wells2022phonetic,choi2024self}) being quantized HuBERT units.
We show that the amount of information in HuBERT units 
can be quantitatively measured,
and a lot of information are in fact present in the discrete units. 
We also show that information is likely to be less complete in the later layers, 
despite more accessible phonetic information,
confirming the choice of WavLM layer \cite{chen2022wavlm} in voice conversions \cite{baas2023voice}.
We reveal that speaker information is sufficiently present in HuBERT discrete units, 
and that phonetic information is sufficiently present in the residual,
showing that vector quantization does not achieve disentanglement.

In our experiments, we empirically evaluate information completeness and accessibility on 
HuBERT representations, along with their discrete units considering different depths of RVQ.
The evaluation on accessibility includes phone classification, pitch estimation and speaker verification.
Our analyses provide insight into the choice of discrete speech units for different speech applications,
and show that information is largely present in the residual.
We remark that 
the discrete units from HuBERT can achieve higher completeness and accessibility if 
we further quantize the residuals, showing better reconstructed log Mels.
\section{Methods}

In the following, we describe the quantization scheme to extract discrete speech units.
We then formally define completeness from an information theory point of view. 
Finally, we draw connections between completeness and accessibility.

\subsection{Discrete speech units with RVQ}

We denote $R$ the speech representations, and $\hat{R}$ the quantized representations after residual vector quantization (RVQ) 
\cite{zeghidour2021soundstream}, also known as multiple stage VQ \cite{juang1982multiple}.
RVQ consists of a cascade of $L$ codebooks, each of which of size $N$, 
successively quantizing the residuals of previous quantization using the nearest
neighbor principle to capture finer details. Different from \cite{zeghidour2021soundstream} that
update the codebooks with exponential moving average, we iteratively 
optimize each codebook using k-means until the loss converges.
Codebooks are not fine-tuned if not specified, following the common practice of discrete speech units derived 
from k-means, where centroids are usually not fine-tuned.
Note that, when $L=1$, our RVQ becomes vanilla k-means.

In practice, we can represent a quantized frame $\hat{r}_t$ 
with discrete speech units $c_t=(c_{t,1},\dots,c_{t,L})$
only at the cost of $L \log_2 N$ bits. 
More formally, let $V=(v_1,\dots,v_L)$ be
the codebooks of RVQ, a quantized frame is
\begin{align}
    \hat{r}_t = \sum_{i=1}^{L} V_i \mathbf{1}_{c_{t,i}}, 
\end{align}
where $\mathbf{1}_{c_i}$ is a one-hot vector with $c_i$-th entry being 1.

\subsection{Completeness as mutual information}

Given (quantized) speech representations, we then define completeness as the mutual information between 
log Mel spectrograms $X$ and the representations.
We choose log Mel spectrograms (log Mels) instead of raw waveforms because 
log Mels are sufficient for many speech processing tasks;
the argument equally applies to waveforms.
We argue that, if a representation is complete, 
it should be able to present \textit{all} information in the log Mel.
The completeness is formally defined as
\begin{equation}
\begin{aligned}
I(R, X) &= H(X) - H(X|R) \\
        &\geq I(\hat{R}, X), 
\end{aligned}
\end{equation}
where the second equation is due to the data processing inequality.
Because $H(X)$ remains constant given different representations, we only have to compute the conditional entropy $H(X|R)$
to measure completeness.

Nonetheless, the desired conditional entropy is generally not available \cite{mcallester2020formal}.
To estimate $H(X|R)$, we upper-bound it with cross entropy estimation,
introducing a variational distribution $q(x|r)$
\cite{mcallester2020formal,pimentel2020information,wolf2023quantifying,liu2024revisiting}.
It leads to a lower bound of mutual information
\begin{align}
    I(R, X) &= H(X) - H(X|R) \notag\\
            &= H(X) + \mathbb{E}_{(x,r) \sim p}[\log q(x|r) + \log \frac{p(x|r)}{q(x|r)}] \notag\\
            &\geq H(X) + \mathbb{E}_{(x, r) \sim p}[\log q(x|r)],
\label{eq:lb}
\end{align}
where $\mathbb{E}_p[\log q(x|r)]$ is the empirical cross entropy,
with the inequality due to the non-negativity of KL divergence.
By making a Gaussian assumption of $q(x|r)$, we obtain
the proposed lower bound
\begin{equation}
\begin{aligned}
    &H(X) + \mathbb{E}_{(x,r) \sim p}[\log q(x|r)] \\%
    &= H(X) - \frac{1}{2} \mathbb{E}_{(x,r) \sim p}[(x - f(r))^2] + \frac{d}{2} \log(2\pi e)  \\%
    &\geq H(X) - \frac{1}{2} \mathbb{E}_{(x,r) \sim p}[(x - f(\hat{r}))^2] + \frac{d}{2} \log(2\pi e),
\label{eq:lb2}
\end{aligned}
\end{equation}
where $d$ is the dimension, $f(\cdot)$ is a regression network, and the third line follows
from the data processing inequality on discrete speech units.
The lower bound implies that,
to achieve a larger lower bound of mutual information, i.e., better estimation of information completeness,
we should minimize the mean square error using a powerful $f(\cdot)$.

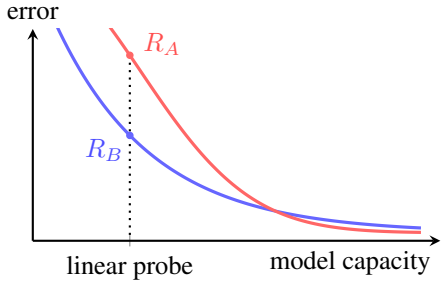
\begin{figure}[htbp]
  \centering
  \begin{tikzpicture}
    \begin{axis}[
      domain=0:2,
      xmin=0,
      xmax=2.1,
      ymin=0,
      ymax=0.8,
      samples=100,
      axis lines=center,
      xlabel={model capacity},
      ylabel={error},
      x label style={anchor=north east},
      y label style={anchor=south},
      xtick={0.5},
      xticklabels={linear probe},
      ytick=\empty,
      height=4.4cm,
      width=7cm,
      enlargelimits=false,
      clip=true,
      thick
      ]
      \addplot [blue!60, very thick, name path = function] {e^(-x*2)+0.03};
      \addplot [red!60, very thick, name path = function] {e^(-x*x*1.6)+0.03};
      \node[inner sep=1pt, label={45:{\textcolor{red!70}{$R_A$}}}] (a) at (50, 65) {};
      \node[inner sep=1pt, label={45:{\textcolor{blue!70}{$R_B$}}}] (a) at (20, 25) {};
      \path[name path=axis] (axis cs:0,0) -- (axis cs:5,0);
      \draw [thick,dotted] (axis cs:0.5, {e^(-0.5^2*1.6)+0.03}) -- (axis cs:0.5,0);
      \node[inner sep=1pt, circle, fill, blue!60] () at (axis cs:0.5, {e^(-0.5*2)+0.03}) {};
      \node[inner sep=1pt, circle, fill, red!60] () at (axis cs:0.5, {e^(-0.5^2*1.6)+0.03}) {};

    \end{axis}
  \end{tikzpicture}
  \caption{An illustration of information accessibility.
    $R_A$ and $R_B$ are two representations, and their
    probing errors differ depending on the model capacity
    of the probes.
    Under a linear probe, information in
    $R_B$ is more accessible than $R_A$ with a lower error.}
  \label{fig:access}
\end{figure}

\subsection{Information completeness and accessibility}

Information accessibility of a representation describes how easy it is
to predict a target information.
Accessibility depends on the capacity of the model used to extract the information,
as shown figuratively in Figure \ref{fig:access}.
A higher model capacity is more likely to have better performance.
To measure information accessibility, previous work has developed various speech downstream tasks 
\cite{oord2018representation,yang2021superb,yang2022autoregressive,lin2023utility}.
It is widely accepted that if a speech property encoded in a representation
is linearly predictable with linear probes (low model capacity),
the information of the speech property in this particular representation is highly accessible \cite{oord2018representation}.
For instance, the information in $R_B$ is more accessible than in $R_A$
with a linear probe in Figure \ref{fig:access}.

On the other hand, information completeness lies at the opposite end of the spectrum, requiring a
higher model capacity to reach a tighter lower bound. 
For example, Figure \ref{fig:access} tells another story if we focus on the high model capacity region.
Similar finding is also noted in \cite{zaiem2023speech}.
To better estimate the mutual information between a speech property and the representations,
as in \eqref{eq:lb},
the cross entropy should be minimized using a powerful $q$. %
This fact is also noted in \cite{voita2020information,pimentel2020information}.
While it is generally not possible to find the optimal $q$ that maximizes the lower bound,
we consider parameterizing $q$ with a deeper network to obtain a tighter lower bound,
treating downstream performance as information accessibility.
\section{Related work}

There are various aspects of literature related to ours.
Given how widely discrete units are applied, especially in speech language models
and voice conversion, we focus on the completeness aspect surrounding
discrete units in this section.

\subsection{Information-theoretic probing}
In this work we focus on information completeness, another aspect of 
a representation, via the lens of information theory.
Several recent approaches have taken information-theoretic 
techniques to evaluate BERT representations \cite{voita2020information,pimentel2020information}. 
Similar techniques have also inspired the evaluation of speech representations \cite{liu2024revisiting},
connecting mutual information to speech downstream tasks.
However, the information completeness of speech representations has not been well studied.

\subsection{Measuring information completeness}
There are other methods claiming that certain speech properties are disentangled in self-supervised speech 
representations \cite{polyak2021speech,qian2022contentvec,lin2023self,zhang2023speechtokenizer} 
or in the extracted discrete units \cite{polyak2021speech,zhang2023speechtokenizer}.
The presence of information is not verified through an information-theoretic measure.
Instead, they take evaluation metrics from voice conversions to measure whether content and speaker information
are preserved in synthesized speech. 
The content and speaker information are analyzed with a speech recognition system 
and a speaker encoder to compare word error rates and speaker similarity between synthesized and original speech
\cite{polyak2021speech,wang2023neural,zhang2023speechtokenizer}.

Although the evaluation protocol is widely adopted,
whether the synthesized speech faithfully reflect the information carried in discrete speech units is questionable.
On one hand, 
the synthesized speech from the representations may hallucinate if the generation involves
GANs \cite{lucas2019adaptive} or diffusion models \cite{aithal2024understanding}.
In GANs, for example, the discriminator only estimates if the generation is real or fake, while not 
estimating the actual distribution \cite{goodfellow2020generative}. 
This prohibits the justification of information completeness on the lower bound of mutual information \eqref{eq:lb}.\footnote{We note that vocoders such as HiFiGAN \cite{kong2020hifi} has a Mel spectrogram loss to 
promote more realistic synthesized speech,
our arguments on the justification of the lower bound still hold.}
In addition to the hallucination on synthesized speech, 
the external speech recognition model can potentially suffer from hallucination as well 
\cite{anguita2005detection,mittal2024towards,frieske2024hallucinations},  
leading to weaker justification of information completeness.
\section{Experimental settings}

Given the lower bound of mutual information \eqref{eq:lb2},
we empirically evaluate the completeness of HuBERT representations,
and the derived discrete units.
We also present accessibility measurements on phonetic classification,
pitch estimation and speaker verification, considering a higher model capacity region.
While one can always argue whether a probing model is sufficiently strong or not,
the aim is \textbf{not} to estimate information content (in fact it is barely possible),
but rather to identify at least how much information is present in the representations.

\subsection{Discrete speech units}
We choose HuBERT layer 4 and layer 9 for all experiments, 
which are the best-performing layers
for content-related and speaker-related tasks respectively \cite{chen2022wavlm}.
Discrete speech units are obtained by running RVQ on these two layers.
We randomly sample 5000 utterances from LibriSpeech \texttt{train-clean-360} \cite{panayotov2015librispeech}
to train each codebook using k-means.
Codebooks are successively optimized to minimize the Euclidean distance between
the quantized and original HuBERT representations.
Unless stated otherwise, RVQ codebooks are not further fine-tuned.
We experiment with $L$ from 1 to 8, denoting $\text{RVQ}_L$
the RVQ with $L$ codebooks.
Note that $\text{RVQ}_1$ is identical to k-means.
Each codebook is of size of 1024, consuming ${\log_2}1024=10$ bits storage cost.

\subsection{Completeness task}
We conduct experiments on information completeness using LibriSpeech.
Models are trained on \texttt{train-clean-360},
and evaluated on \texttt{dev-clean}. The sampling rate is 16000.
We use 80 bands log Mels as $X$ in \eqref{eq:lb2}, the target of completeness. To match the frame rate of 
HuBERT representations (50 Hz), we set a hop size of 320. We set the length of the FFT to 1024.
We do not normalization log Mels with global mean and variance.
We parameterize $f$ as convolutional networks. It consists of two parts. 
The first part contains 6 convolutions with channels $(256, 256, 256, 256, 512, 512)$,
strides $(1, 1, 1, 1, 2, 2)$ and a kernel size of 3.
The second part contains 8 ConvNeXt blocks used in \cite{siuzdak2023vocos}.
We use a batch size of 16 and a learning rate of 0.0002. Models are trained for up to 60 epochs.
The objective is to predict log Mels by minimizing MSE, which equivalently maximizes 
the lower bound of mutual information.

\subsection{Accessibility tasks}
We design three tasks to evaluate information accessibility, taking into account
probes with higher model capacity as oppose to \cite{lin2023utility}.
We choose phone classification (PC) to evaluate the presence of phonetic information.
In particular, discrete speech units have shown strong phonetic prominence in \cite{wells2022phonetic}.
To see if prosody information is preserved in discrete units, 
we conduct experiments on pitch estimation ($f_0$).
For the third task, we present speaker verification (SV) to measure if speaker-related information in present.

We use 3-layer feedforward networks following by a linear layer
to model phone classification and pitch estimation on Wall Street Journal (WSJ) \cite{paul1992design}.
We set the hidden dimension to 3076 and use ReLU as the activation function for each 
feedforward network.
We adopt the setups in \cite{yang2022autoregressive} and train models on the WSJ training
set using 90\% of \texttt{si284}. We select the best model based on its performance on
the development set, the rest 10\% of \texttt{si284}. 
We report numbers on \texttt{eval92} after training.
Models are trained with a learning rate of $0.001$ using a batch size of $12$.

For phone classification, we use forced alignments extracted by a speaker adaptive GMM-HMM 
as targets. Its performance is measured using phone error rates (PER).
Regarding pitch estimation, we extract the fundamental frequency using PYIN
\cite{mauch2014pyin}, and treat them as the ground truth. 
The minimum and maximum frequency in set to be 50 Hz and 600 Hz respectively.
We use root-mean-square error (RMSE) in Hz and predict pitch only on sonorants obtained from
the forced alignments.

Finally, we evaluate whether representations and discrete speech units encode speaker information
by performing speaker verification (SV) on voxceleb1 \cite{nagrani2017voxceleb}.
We employ a variant of ECAPA-TDNN \cite{desplanques2020ecapa} to learn speaker embeddings,
where we do not concatenate the mean and standard deviation before attentive pooling.
We also do not use AAM-Softmax as in the original paper.
Speaker encoders are trained with a learning rate of $0.0005$ using a batch size of $8$. 
We crop the input utterance to at most 12 seconds due to the memory constraint.
We train all models for 10 epochs.

\begin{figure}[htp]
    \centering
    \includegraphics[width=12em]{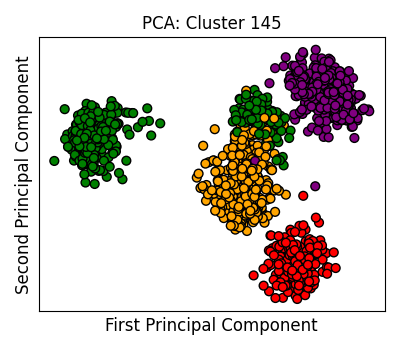}
    \includegraphics[width=12em]{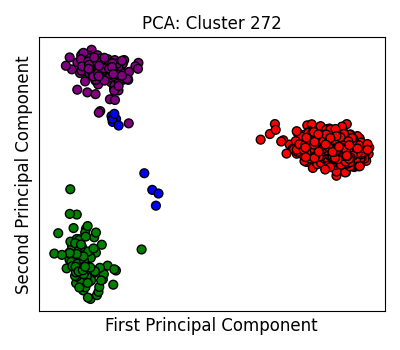}
    \caption{\small Frames of HuBERT representations assigned
      to two example k-means clusters are visualized with
      the first two principle components of PCA.
      Colors represent speaker identifies.}
    \label{fig:cluster}
\end{figure}
\begin{table}[htp]
\centering
\captionof{table}{Accessibility of phone identifies, $f_0$, and speaker identities
  on the 4th and the 9th HuBERT layer. Residuals are computed 
  by subtracting the centroids from the associated representations.}
\resizebox{0.93\columnwidth}{!}{%
\begin{tabular}{c|ccc}
\toprule
                                & PC       & $f_0$   & SV     \\ 
                                & PER (\%)       & RMSE (Hz)   & EER (\%)     \\\midrule
HuBERT L4                       & 11.6      & 35.7   & 4.4     \\ 
k-means ($\text{RVQ}_1$)        & 29.8      & 67.1   & 18.8    \\
Residual                        & 13.2      & 37.8   & 5.5     \\ \midrule
HuBERT L9                       & 7.3       & 41.0   & 6.5     \\
k-means ($\text{RVQ}_1$)        & 23.4      & 72.9   & 21.5    \\
Residual                        & 8.3       & 41.7   & 7.3     \\ \bottomrule
\end{tabular}
\label{tab:residual}
}
\end{table}

\section{Results and Discussions}

\subsection{Information in the residuals}
We first show evidence of the presence of speaker information in the 
residuals of k-means ($\text{RVQ}_1$).
We randomly pick 300 utterances from 6 speakers in LibriSpeech \texttt{dev-clean},
and present HuBERT 9th layer frames assigned to two sample clusters.
Frames belong the same speaker are in the same color. 
Representations assigned to the two clusters are shown in Figure \ref{fig:cluster} using PCA,
with the presence of speaker's information.
Similar patterns are also observed in several k-means clusters, as noted in \cite{lin2023self}.
The evidence indicates that information in the residuals should be further mined by increasing
the cluster size or more efficiently RVQ.

\begin{table*}[t]
\begin{center}
\caption{Results of information completeness and information accessibility. 
    The information rate, known as the storage cost per frame is also included. 
    A lower MSE means the representations
    are closer to complete.
\label{tab:1}
}
\resizebox{1.6\columnwidth}{!}{%
\setlength{\tabcolsep}{10pt}
\begin{tabular}{l|cccccc}
\toprule
                   & \multicolumn{2}{c}{\makebox[0pt]{Information completeness}}    & \multicolumn{3}{c}{Information accessibility}  & Information rate  \\ 
                   & MSE $\downarrow$    & SNR (dB) $\uparrow$ & PER (\%) $\downarrow$  & RMSE (Hz) $\downarrow$   & EER ($\%$) $\downarrow$  & Per frame bits   \\ \midrule
Log Mel            & 0.0     & $\inf$     & 37.3      & 38.4    & 13.2                                               & $32 \times 80$ \\\midrule
HuBERT L4           & 39.6   & 18.5     & 11.6      & 35.7   & 4.4                                             & $32\times 768$\\             
$\text{RVQ}_8$(fine-tuned)  & 49.5  & 17.5    & 13.8 & 49.8      & 5.9                                               & $10\times 8$  \\ 
$\text{RVQ}_8$     & 63.8    & 16.4     & 22.5      & 60.7   & 9.9                                            & $10\times 8$  \\
k-means             & 82.6   & 15.3     & 29.8      & 67.1   & 18.8                                            & $10\times 1$ \\\midrule
HuBERT L9          & 54.2    & 17.1      & 7.3       & 41.0   & 6.5                                             & $32 \times 768$\\
$\text{RVQ}_8$     & 75.1    & 15.7     & 14.8      & 64.7   & 12.8                                            & $10\times 8$  \\
k-means             & 91.6   & 14.8     & 23.3      & 72.9   & 21.5                                            & $10\times 1$  \\ \bottomrule
\end{tabular}
}
\label{tab:compare}
\end{center}
\vspace{-1em}
\end{table*}

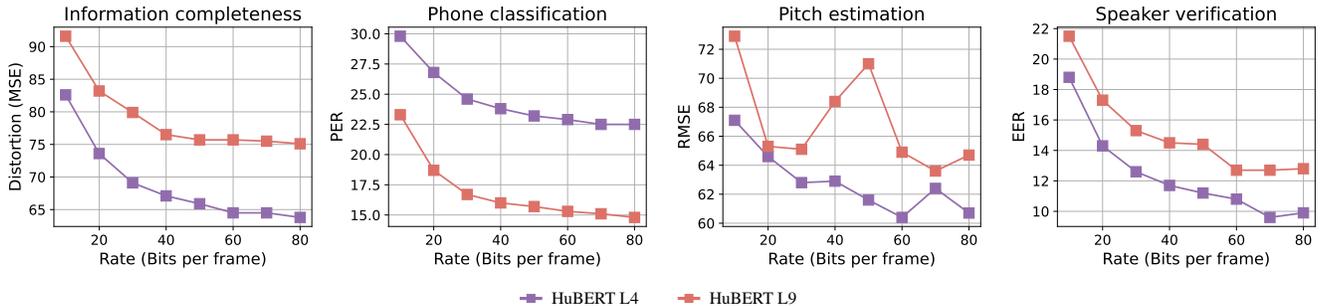
\begin{figure*}[htp]
    \centering
    \includegraphics[width=12.4em]{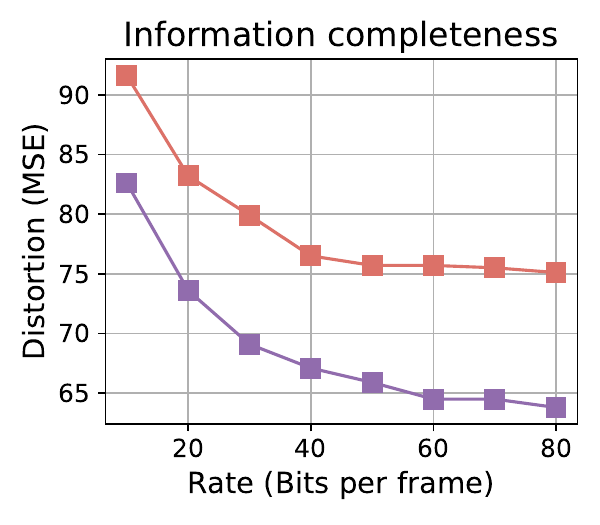}
    \includegraphics[width=12.4em]{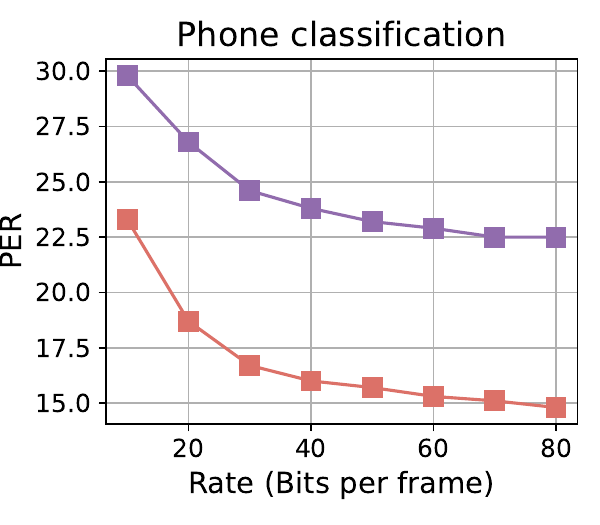}
    \includegraphics[width=12.4em]{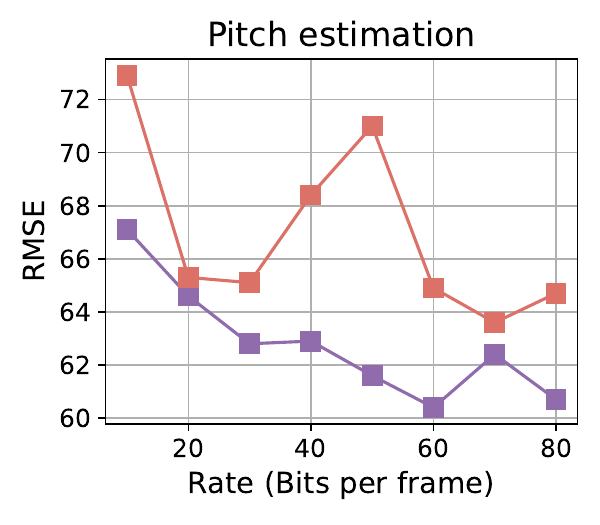}
    \includegraphics[width=12.4em]{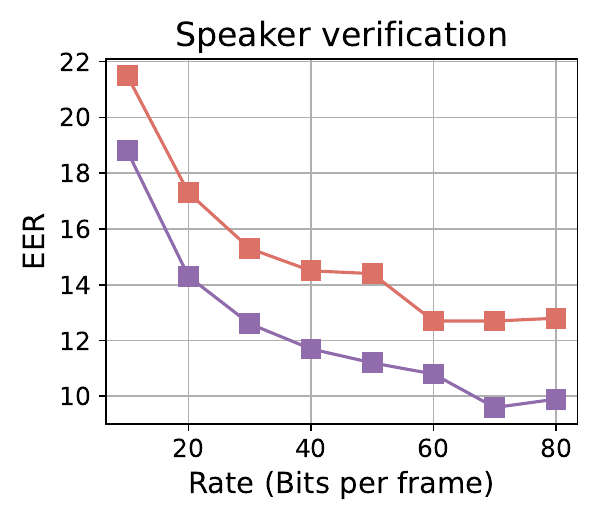}
	\begin{tikzpicture}[font=\footnotesize]
	    \draw[very thick, c4] (0, 3.5) -- (0.3, 3.5);
        \draw[c4, mark=square*, mark size=1.5pt] plot coordinates {(0.15, 3.5)};
		\node[anchor=west] at (0.3, 3.5) [font=\fontsize{6.5}{0}\selectfont] {HuBERT L4};
	    \draw[very thick, c9] (2.1, 3.5) -- (2.4, 3.5);
        \draw[c9, mark=square*, mark size=1.5pt] plot coordinates {(2.25, 3.5)};
		\node[anchor=west] at (2.4, 3.5) [font=\fontsize{6.5}{0}\selectfont] {HuBERT L9};
	\end{tikzpicture}
    \caption{The completeness and accessibility of representations at different
    rates (bits per frame). We vary the depth of RVQ from $L=1$ to $L=8$.
    Representations are quantized at a cost of 10 bits per codebook, 
    corresponding to a codebook size $N=1024$. Codebooks are not fine-tuned.}
    \label{fig:rd}
    \vspace{-1em}
\end{figure*}

\subsection{Information disentanglement?}
Previous work has claimed the disentanglement properties of self-supervised representations and their
discrete units after k-means \cite{polyak2021speech,zhang2023speechtokenizer}.
To test the claim, we evaluate the original representations,
including HuBERT 4th layer (HuBERT L4) and HuBERT 9th layer (HuBERT L9), 
their k-means ($\text{RVQ}_1$) units and their residuals ($R - \hat{R}$) after k-means.
We want to emphasize that the gap between the original representations and the residuals does not
imply information loss after quantization as we cannot tell how tight the lower bound of mutual information is.
Nonetheless, we can verify whether the information is present or even disentangled.

Table \ref{tab:residual} reports the results on the accessibility tasks.
We first note that speaker and phonetic information is sufficiently present in HuBERT discrete units on both layers. 
The strong performance on the residuals indicates that information 
remains present after vector quantization.
We observe little disentanglement of the speech properties but the likelihood of information loss in general.
While claiming information loss is theoretically difficult,
we do find it hard to recover performance even with stronger probes.
We also observe that pitch is less accessible with k-means units of HuBERT L9.

\begin{figure*}[htp]
    \centering
    \begin{adjustbox}{scale=1.13}
    \begin{subfigure}{.5\columnwidth}
    \includegraphics[width=\columnwidth]{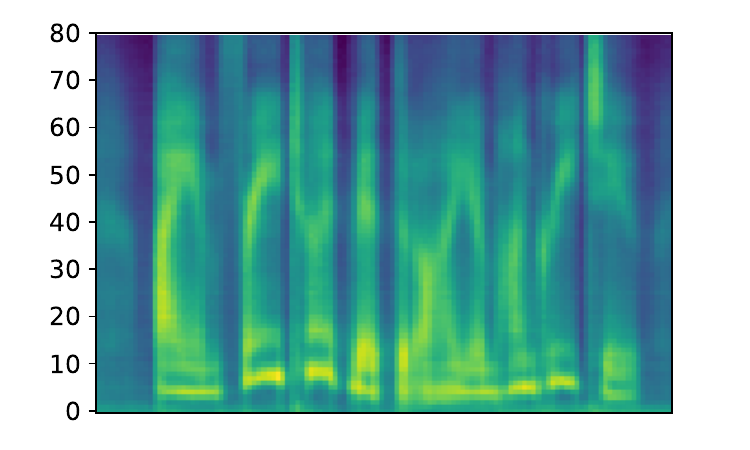}
    \caption{k-means}
    \end{subfigure}\hspace{-5mm}
    \begin{subfigure}{.5\columnwidth}
    \includegraphics[width=\columnwidth]{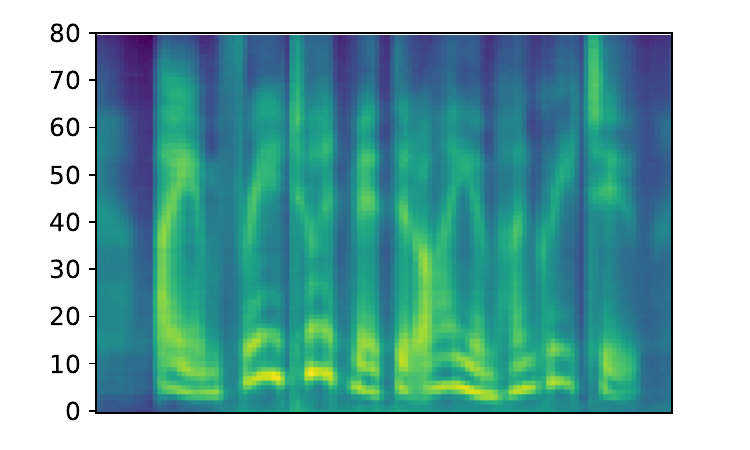}
    \caption{$\text{RVQ}_8$}
    \end{subfigure}\hspace{-5mm}
    \begin{subfigure}{.5\columnwidth}
    \includegraphics[width=\columnwidth]{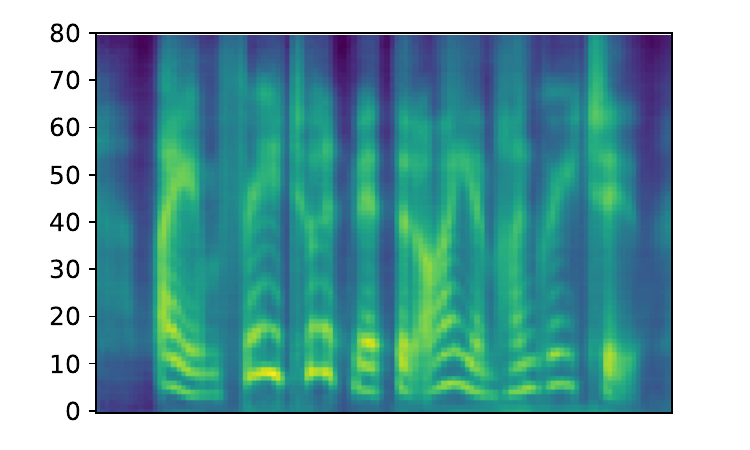}
    \caption{$\text{RVQ}_8$ (fine-tuned)}
    \end{subfigure}\hspace{-5mm}
    \begin{subfigure}{.5\columnwidth}
    \includegraphics[width=\columnwidth]{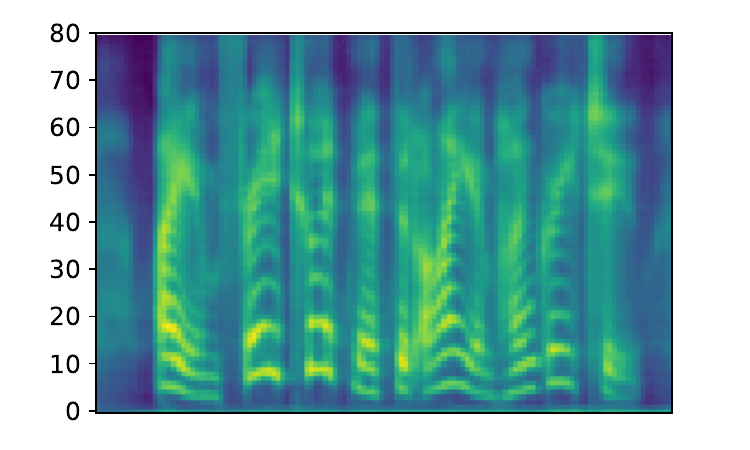}
    \caption{HuBERT L4}
    \end{subfigure}\hspace{-5mm}
    \end{adjustbox}
    \caption{An example of the reconstructed log Mels with HuBERT L4 representations and their discrete units.
        The distortion (MSE) decreases from left to right. Details over 20 Mel bands are better captured in (c)
        and (d). The ground truth is shown in Figure \ref{fig:mel-gt}.
    }
    \label{fig:mels}
    \vspace{-1em}
\end{figure*}

\begin{figure}[htp]
  \begin{minipage}{0.6\columnwidth}
    \centering
    \includegraphics[width=14.6em]{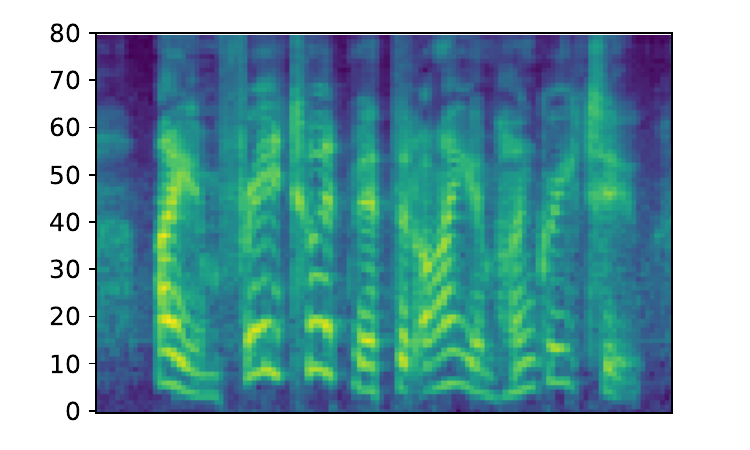}
    \caption{The ground truth utterance for Figure \ref{fig:mels}.}
    \label{fig:mel-gt}
  \end{minipage}\hfill
  \begin{minipage}{0.4\columnwidth}
    \centering
    \captionof{table}{The completeness of the 4th, 9th and 12th HuBERT layer.}
    \begin{tabular}{ll}
    \toprule
    HuBERT          & MSE        \\\midrule
    L4              & 39.6       \\
    L9              & 54.2        \\
    L12             & 52.8       \\\bottomrule
    \end{tabular}
    \label{tab:layer}
  \end{minipage}\hfill
\end{figure}

\subsection{Information completeness and accessibility}

We have revealed in the previous sections that residuals contains much information that should not be discarded.
We conduct RVQ for up to 8 codebooks to capture the information in the residuals.
Codebooks in RVQ are optimized with iterative k-means, and hold fixed unless otherwise stated.
The number of bits used to encode single frame is $L {\log_2}N = L \times 10$. 
Here, we explore how complete and accessible the information encoded in the discrete speech units.

Table \ref{tab:compare} summarizes the completeness and accessibility of representations before and after 
vector quantization.
Besides our completeness objective MSE, we provide signal-to-noise ratio (SNR) in dB
to gain intuition of the reconstruction quality.
Rate represents the number of bits per frame to be stored or transmitted in speech coding.
We provide log Mels as the upper bound of completeness. 
Despite the most complete baseline compared to 
other representations, the phonetic information encoded in log Mels is less accessible 
than HuBERT representations by a large margin in phone classification, even their discrete units.
Compare to HuBERT L9, L4 is closer to complete, 
showing better performance in pitch estimation and speaker verification.
On the other hand, HuBERT L9 exhibits higher phone accessibility, outperforming log Mels with the rate of 10 bits.

The results provide a detailed assessment of speech representations and discrete speech units. For example,
HuBERT L4 is more preferred than L9 in voice conversion \cite{baas2023voice}, 
speech codecs \cite{zeghidour2021soundstream,defossez2022high}
and discrete units for speech language modeling \cite{lakhotia2021generative,borsos2023audiolm,wang2023neural}.
The lower bound of mutual information can also be used to quantify 
the redundancy between two signals \cite{wolf2023quantifying}. 
Based on our results, the claim made in \cite{zhang2023speechtokenizer} about the 
significant redundancy between HuBERT units and speech properties
is not about whether the units are semantic or disentangled
but likely due to information loss or the lack of model capacity.
In fact, HuBERT units adequately capture information in acoustic features.

\subsection{Fine-tuning RVQ on the lower bound of MI}
The proposed lower bound can not only be used to measure information completeness but also improve the learned discrete units.
We experiment with fine-tuning the codebooks of $\text{RVQ}_8$ by maximizing the lower bound \eqref{eq:lb2}
with convolution networks $f$, denote $\text{RVQ}_8$ (fine-tuned).
We only fine-tune the codebooks once with log Mels.
Unlike in SoundStream \cite{zeghidour2021soundstream} 
that updates codebooks with exponential moving average, 
we simply use Gumbel Softmax with a constant temperature of 1 \cite{jang2016categorical,maddison2016concrete}.
Quantizer dropout is not applied.
We find that fine-tuning the codebooks results in an increase in completeness and accessibility of all tasks 
for $\text{RVQ}_8$ (fine-tuned).
Moreover, it outperforms HuBERT L9 in completeness and speaker verification with 80 bits storage.

\subsection{Rate-distortion and rate-accessibility}
We carry out experiments to study the effects of RVQ depth on information completeness and accessibility,
showing the importance of mining residuals and its trade-off between the compression rate and the performance.
Figure \ref{fig:rd} shows the rate-distortion and rate-accessibility curves from 10 bits ($L=1$) 
to 80 bits ($L=8$).
As expected, increasing $L$ generally improves information completeness and accessibility. The variations in pitch estimation 
is relatively large in HuBERT L9 before 60 bits.
The trade-off between rate and distortion is important for deciding the
information processing capabilities of representations for different applications.

The distortion is reflected in the predicted log Mels as shown in Figure \ref{fig:mels}.
Discrete units with k-means struggle to capture the first two harmonics, while $\text{RVQ}_8$ starts to 
capture the rises and falls of the first three harmonics.
By fine-tuning codebooks on the lower bound of completeness,
$\text{RVQ}_8$ (fine-tuned) predicts clearer spectrograms only at a cost of 80 bits. We present the ground truth in
Figure \ref{fig:mel-gt} as a reference.

\subsection{Information in the last layer}
With the lower bound, it is also interesting to see how much information is preserved 
in the last layer.
As shown in Table~\ref{tab:layer}, HuBERT L12 achieves a larger lower bound (lower MSE) than L9.
Due to the data processing inequality, this implies that L9 is at least as complete as HuBERT L12.
Similarly, the first three layers is at least as complete as layer 4.

\section{Conclusion}
We present an information-theoretic approach to
estimating the completeness of speech representations before and after vector quantization.
In addition, we establish connections between information completeness and information accessibility,
providing a lower bound of completeness with a stronger justification.
We then use the concepts of completeness and accessibility to validate
claims on the information encoded 
in HuBERT representations, including the disentanglement and 
the redundancy of discrete units.

We further explore the relationships among information completeness, accessibility and rate,
showing the trade-off between depths of residual vector quantizer (the rate) and the other two quantities.
Our results re-position the role of self-supervised discrete units on speech applications,
showing that in addition to phonetic information,
prosody and speaker information can also be captured by quantizing the residuals.

\bibliographystyle{IEEEbib}
{\small
\bibliography{main}}

\end{document}